# HOW DECENTRALIZED IS THE GOVERNANCE OF BLOCKCHAIN-BASED FINANCE?

Empirical Evidence from four Governance Token Distributions

*Research in Progress*

**Johannes Rude Jensen**
University of Copenhagen
eToroX Labs
johannesrudejensen@gmail.com

**Victor von Wachter**
University of Copenhagen
victor.vonwachter@di.ku.dk

**Omri Ross**
University of Copenhagen
eToroX Labs
omri@di.ku.dk

## Abstract

*Novel blockchain technology provides the infrastructure layer for the creation of decentralized applications. A rapidly growing ecosystem of applications is built around financial services, commonly referred to as decentralized finance. Whereas the intangible concept of 'decentralization' is presented as a key driver for the applications, defining and measuring decentralization is multifaceted. This paper provides a framework to quantify decentralization of governance power among blockchain applications. Governance of the applications is increasingly important and requires striking a balance between broad distribution, fostering user activity, and financial incentives. Therefore, we aggregate, parse, and analyze empirical data of four finance applications calculating coefficients for the statistical dispersion of the governance token distribution. The gauges potentially support IS scholars for an objective evaluation of the capabilities and limitations of token governance and for fast iteration in design-driven governance mechanisms.*

*Keywords: Distributed Ledger Technology, Blockchain, Decentralized Finance (DeFi), Governance, Governance Token*

# 1      Introduction

In recent years, blockchain technology has been of significant interest to scholars in the strategic information systems (IS) genre (Lindman, Tuunainen and Rossi, 2017; Rossi *et al.*, 2019; Kolb *et al.*, 2020). In the academic and practitioner literature alike, the abstract concept of 'decentralization' is typically presented as a key value driver for applications implemented on permissionless blockchain technology (Zheng *et al.*, 2017; Treiblmaier, 2019). Decentralization as a design objective largely aims to provide open and resistant protocols, reducing dependencies on centralized agency (Zheng *et al.*, 2017). Yet, results from multiple independent studies has exposed a striking tendency for the concentration of assets in the largest permissionless blockchain networks Bitcoin and Ethereum (Böhme *et al.*, 2015; Azouvi, Maller and Meiklejohn, 2018; Wu *et al.*, 2020).

The entrepreneurs designing the latest generation of blockchain-based decentralized financial applications, colloquially referred to as 'DeFi', seek to accomplish a 'decentralized' distribution of voting power amongst network participants through the issuance of governance tokens; fungible entities enabling holders to participate directly in decision-making processes through majority voting schemes. Governance tokens trade on secondary markets and thus affords team members and early stakeholders the opportunity to raise funding through the capital formation associated with the distribution of tokens (Kranz, Nagel and Yoo, 2019).

Thus, methodologies for disseminating governance tokens attempt to strike a balance between the relative 'decentralization' of voting power amongst a wide span of active or passive stakeholders while simultaneously incentivizing application usage and securing funding for the core team. In this paper, we present ongoing efforts towards a unified framework for the evaluation of governance token distributions. We approach the research question: *How decentralized is the governance token distribution in DeFi applications?* In the absence of a standardized quantifiable definition of the abstract concept 'decentralization', we draw on the work of (Srinivasan and Lee, 2017) in measuring the statistical dispersion of governance tokens by computing the Gini- and Nakamoto-coefficients for the distributions.

The Gini-coefficient represents the statistical dispersion of assets or income over a large sample indicating a measurement of equality. The Nakamoto-coefficient represents the minimum number of entities whose cumulative proportions sum to a 51% stake, effectively denoting the number of colluding entities required to receive binary majority. We collect, parse and analyse data from four recent governance token distributions associated with reputable DeFi applications: Balancer[1], Compound[2], Uniswap[3], Yearn Finance[4]. At the time of writing, these applications hold an aggregate $5.081 billion of assets on the Ethereum blockchain, with a combined valuation at $1.692 billion[5]. As follows, the distribution of governance tokens is an increasingly important issue, as ownership of these tokens defines the voting rights for the novel DeFi applications. By collecting and analysing empirical data on the token economies emerging around decentralized financial applications, we contribute to the growing IS discourse on the capacity for permissionless blockchain technology to enact socio-economic change through on-chain governance.

---

[1] More information on the specific application can be found at https://www.balancer.finance

[2] More information on the specific application can be found at https://www.compound.finance

[3] More information on the specific application can be found at https://www.uniswap.org

[4] More information on the specific application can be found at https://www.yearn.finance

[5] Data from https://coingecko.com on Oct. 31$^{st}$, 2020

## 2 Blockchain Technology and Decentralization

Blockchain technology denotes a type of a distributed database architecture maintaining a shared state. Later iterations of the technology have introduced a deterministic execution environment with a higher-level programming language capable of executing small Turing-complete programs, commonly referred to as 'smart contracts ' (Swan, 2015; Antonopoulos and Wood, 2018). Smart contracts enable decentralized applications running on top of the blockchain, inheriting the core property of *decentralization* (Zheng *et al.*, 2017; Treiblmaier, 2019). The popularization of smart contract infrastructure has brought forth a new form of organizational structure, colloquially referred to as 'decentralized autonomous organizations' (DAOs). DAOs are governed by deterministic rules written with smart contracts, which facilitates the coordination between unknowing agents in a trust minimized setting (Wright and De Filippi, 2015). Typically, decentralized applications start centralized allowing the developers to iterate core functionality and push upgrades in an instant. Following the initial phase, these applications largely seek to accomplish the 'decentralization' of governance through establishment of a robust DAO fostering participation of active stakeholder.

Building on top of the infrastructure layer, decentralized applications are subject to the underlying blockchain. As such a decentralized infrastructure is a prerequisite for a decentralized application. Several scholars contributed studies analysing the largest permissionless blockchain networks Bitcoin and Ethereum (Böhme *et al.*, 2015; Srinivasan and Lee, 2017; Azouvi, Maller and Meiklejohn, 2018; Wu *et al.*, 2020). Drawing from (Wu *et al.*, 2020) 4 mining pools control the majority (50.5%) of the compute power securing the Bitcoin network and 3 pools control the majority (62.3%) of the compute power securing the Ethereum network. Whereas the richest 10 addresses hold 17.67% of the value on the Bitcoin network, the richest 10 addresses hold 12.02% of the value on the Ethereum network. The concentration of compute power amongst a declining number of entities has been shown to be the product of 'mining-pools', in which multiple operators syndicate to pool computational resources with the intention of increasing the aggregate output whilst fixing income for members. The concentration of wealth appears to be a result of two primary factors (I) the tendency for capital accumulation amongst wealthy holders and, to a larger degree (II) the initial distribution of native assets amongst a small selection of initial stakeholders. As the market capitalization of the network increases with the influx of new participants and subsequent demand, the value of the native asset or governance token appreciates, often resulting in a skewed distribution of wealth. This tendency presents a stark contradiction to the original ethos of permissionless blockchain technology: The decentralization of governance and voting power amongst multiple, non-colluding, agents.

### 2.1 Governance Tokens

Governance tokens are fungible units implementing a voting logic amongst a set of stakeholders by which holders can express their intention for the protocol development in majority-voting schemes. Stakeholders express their opinion in a variety of protocol specific voting mechanisms, in which an account balance in the governance token is used in signalling for or against a proposal in a binary voting scheme. As such, it follows that the relative distribution of governance tokens amongst network participants is an expression of the degree of 'decentralization' of the protocol. Like traditional equities, governance tokens are fungible entities and trade on secondary markets which facilitates price discovery and capital formation. New distribution schemes for governance tokens address the issue of asset concentration through a diverse set of distribution methodologies, typically with the intention of incentivizing platform throughput volume or liquidity, through so-called 'yield-farming' schemes.

Currently a rapidly growing ecosystem of blockchain-based applications is being built around financial services. The surge of popularity saw the aggregate value of assets under management across DeFi applications grow from a range of $400-500 million at the outset of 2020 to an excess of $13.6 billion[6] at the end of October of the same year. As follows, the distribution of governance tokens is an increasingly important issue, as ownership of these tokens defines the voting rights for smart contract-based platforms holding several billion dollars and occasionally processing more transaction volumes than the leading centralized orderbook exchanges.

The governance token distribution methodologies differ widely, as teams evaluate the balance between decentralization, incentivized participation and the need to secure venture capital (Kranz, Nagel and Yoo, 2019). In Table 1, we present the distribution methodologies for the governance tokens of the four selected projects Balancer, Compound, Uniswap and Yearn Finance.

| Application | Initial Allocation | | | | |
| --- | --- | --- | --- | --- | --- |
| | Retrospective Users | Participation Incentives | Founders & Team | Investor & Advisor | Ecosystem Treasury |
| BAL | 0% | 65% | 5% | 25% | 5% |
| COMP | 0% | 42.37% | 26.05% | 23.76% | 7.82% |
| UNI | 15% | 45% | 21.82% | 18.18% | 0% |
| YFI | 0% | 100% | 0% | 0% | 0% |

*Table 1.  The initial allocation of governance tokens by project*

The distribution of governance tokens is typically directed towards internal parties as compensation for their work ('Founder & Team') or capital ('Investor & Advisor'). External agents can be incentivized for future application usage ('Participation Incentives') promoting user adoption or long-term ecosystem development ('Ecosystem Treasury'). Finally, some project issue governance tokens to the earliest of all users having contributed to the application ('Retrospective Users').

## 3 Introducing the Dataset

We aggregate, parse and clean data from the Ethereum blockchain for the following four DeFi projects: Balancer, Compound, Uniswap and Yearn Finance. The projects were selected by their relative maturity and significant volumes processed. The dataset can be retrieved publicly on any Ethereum node. In Table 2 we present an overview of the four applications as of October 31st 2020.

| Application | Token | Tokens created (Fully diluted) | Number of addresses | On-chain governance deployment |
| --- | --- | --- | --- | --- |
| Balancer | BAL | 38,335,000 (100,000,000) | 21,362 | Jun-01-2020 |
| Compound | COMP | 10,000,000 (10,000,000) | 64,823 | Jun-10-2020 |
| Uniswap | UNI | 911,031,617 (1,000,000,000) | 98,100 | Sep-14-2020 |
| Yearn Finance | YFI | 30,000 (30,000) | 12,766 | Jul-30-2020 |

---

[6] Data from https://defipulse.com on Oct. 31st 2020

*Table 2.       Overview of the initial dataset*

Diligence and precision in the cleaning and evaluation of data is of the utmost importance when aggregating data from public blockchains. Since addresses are generated using public-key cryptography, entities behind addresses are pseudonymous. Thus, due to the pseudonymous nature of blockchain technology determining if an agent holds multiple addresses is not possible. In this work, we make several implicit assumptions with potential implications for the integrity of the results. Based on best practices, we applied minimal intervention, reducing the cleaning to two principles: Prune the addresses which technically cannot participate in governance to date as well as addresses with dust values, where participating in governance is economically not feasible. Table 3 elaborates on the address types and reasons for exemption.

| Address type | Description | Reasons for exemption |
|---|---|---|
| Timelocked address | The tokens of this address cannot be used for governance as they are locked until a specific date. For exemption the date must have been disclosed publicly. | Tokens cannot participate in governance if they are locked. |
| Dead address | The address cannot be accessed. | There is no private key to sign transactions. |
| Pool address | The address is known to be used by multiple entities. For example, addresses administered by exchanges. | This is comparable to the real world, where a bank pools the funds of arbitrary many customers. We assume the bank will remain neutral with customer funds. |
| Smart contract address | Smart contract address without logic to govern the ownership of tokens. Not an externally owned address. | These smart contracts are created for executing tasks and do not specifically implement logic to govern tokens. |
| Low balance address | The costs of the transaction to vote is higher than the economic value of the voting power. | Participation in voting is economically not feasible. |

*Table 3.       Technical and economic reasons to exclude governance tokens from voting*

Examining contract code repositories alongside public investor, advisor and team relations for the four distributions, we identified and removed a total of 366 addresses controlled by non-participatory entities (Table 4). In several cases, the addresses removed held vast amounts of governance tokens, either vested, retained, or otherwise removed from circulation.

| App | Number of addresses | Remaining addresses | Tokens held by all addresses | Tokens held by the remaining addresses |
|---|---|---|---|---|
| BAL | 21,362 | 21,276 (-86) | 38,335,000 | 3,688,812 |
| COMP | 64,823 | 64,691 (-132) | 10,000,000 | 3,945,760 |
| UNI | 98,100 | 97,980 (-120) | 911,031,617 | 88,801,074 |
| YFI | 12,766 | 12,738 (-28) | 30,000 | 17,395 |

*Table 4.       Governance token analyzed after data preparation*

## 4 Analysis

We calculate the Gini coefficient and Nakamoto coefficient for the governance token distribution of each application. Sorting the list of addresses in ascending order such that x has the rank i, we compute the Gini-coefficient G, as:

$$G = \frac{\sum_{i=1}^{n}(2i - n - 1)x_i}{n \sum_{i=1}^{n} x_i}$$

Let x be an observed value, n the number of values observed and i the rank of values in the ascending order. We interpret the Gini coefficient as measurement of inequality, indicating the list of addresses proximity to a uniform distribution. Following (Srinivasan and Lee, 2017) we compute the Nakamoto-coefficient N for a distribution d with K entities in which addresses $a_1 \ldots a_K$ is the addresses controlled by each of the K entities operating on the network, defined as:

$$N_d := min\left\{ \in [1, \ldots, K] : \sum_{i=1}^{k} a_i \geq 0.51 \right\}$$

As follows, the Nakamoto coefficient for a distribution d is the smallest number of entities whose propositions sum to >51% of governance tokens in circulation. In figure 1 we present the results of the analysis, plotting the Lorenz curve in green. The area below perfect equality and Lorenz curve is the Gini coefficient. The Nakamoto coefficient is indicated by the red, dotted line.

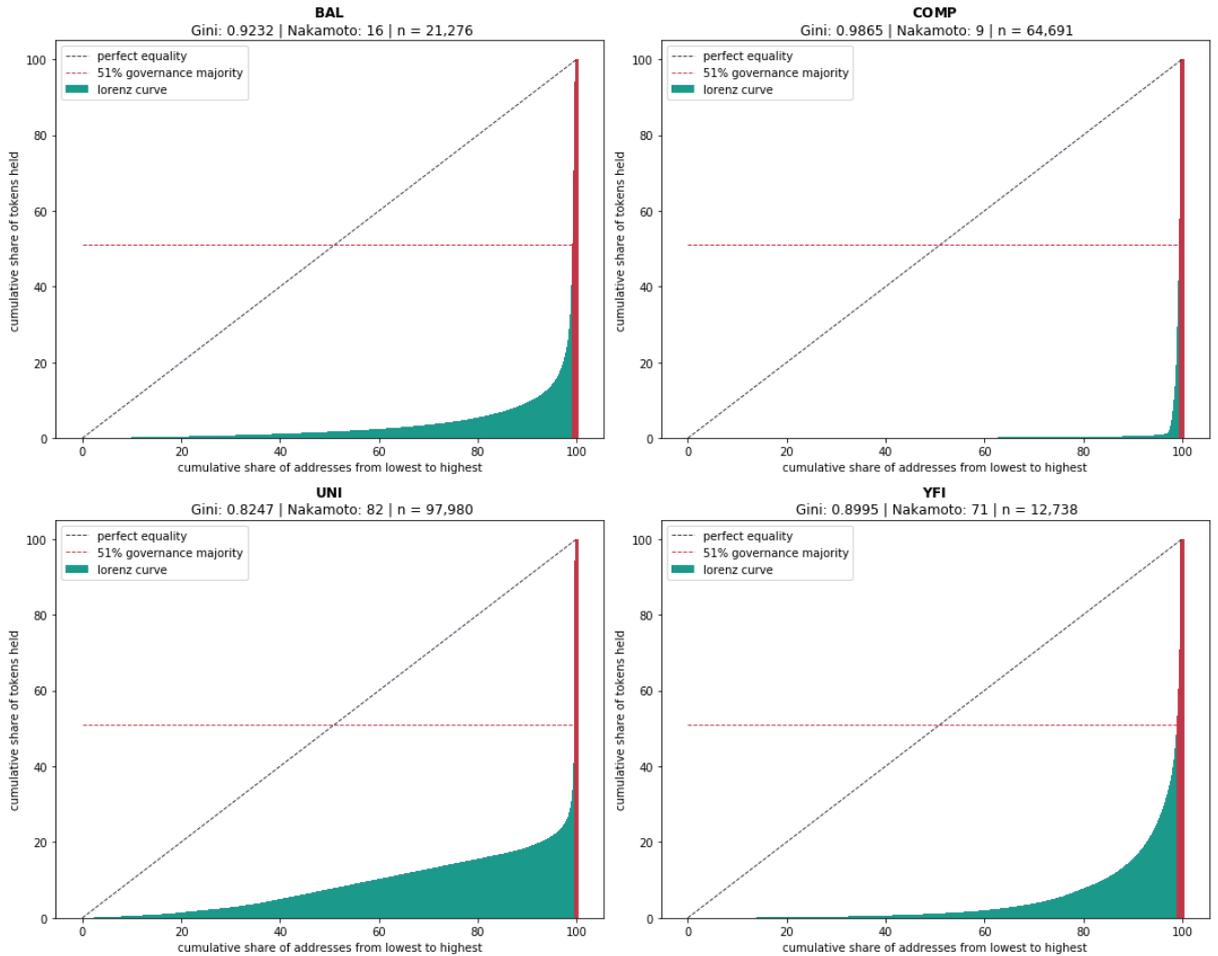

*Figure 1.     Gini and Nakamoto coefficients for the four governance tokens*

Summarizing the results, we show the voting power controlled by the richest 5, 100 and 1000 addresses. As evident in Figure 2, none of the distribution methodologies successfully achieved a Nakamoto coefficient surpassing 100 addresses, indicating that none of the projects require a quorum of more than a hundred addresses to effectively enact governance decisions.

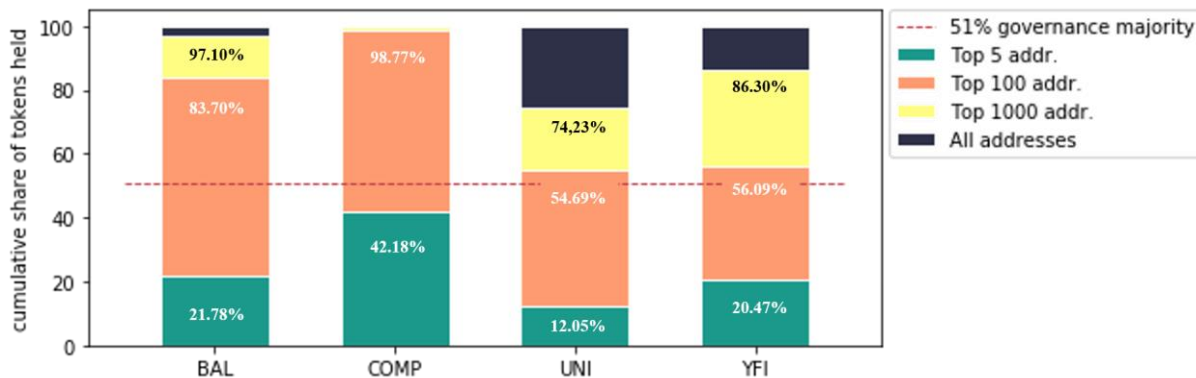

*Figure 2.         Governance token distribution amongst top addresses*

## 5      Discussion

The differing distribution methodologies summarized in Table 1 are widely reflected in the measures of decentralization. Our results indicate that the unique retrospective distribution of UNI has proved moderately successful, generating the lowest Gini and highest Nakamoto coefficient for all distributions, thus approximating the highest degree of 'decentralization'.

The somewhat unconventional distribution of YFI in which all tokens were allocated proportionally as stakeholder incentives for application usage has resulted in a reasonable degree of decentralization, albeit favouring wealthy stakeholders. Comparatively, the initial distribution of COMP may be considered suboptimal, as voting power remains largely concentrated. Only 42,37% of the governance tokens was distributed through incentivized participation. Unsurprisingly, the involvement of early-stage venture capital investors appears to correlate with a higher concentration of governance tokens amongst fewer addresses.

Token voting is a promising first step towards transparent, open, socio-economic governance. The governance of DeFi application largely depends on the governance token distribution and protocol specific voting mechanism. Due to the novelty of blockchain applications, we expect that entrepreneurs will iterate and try different instantiations of on-chain governance. However, not too dissimilar from the early development of the internet, it appears that utilizing decentralized infrastructure like blockchain technology does not necessarily lead to a decentralization of authority of the application layer. In comparison the economy of scale and the value of data led to a relative concentration of services building on top of the internet. Furthermore, we expect that flawed governance design potentially makes it susceptible to both large stakeholders controlling the protocols or hostile exploits. Attack vectors are multifaceted and have been exploited in the past (Zoltu, 2019). The need for proper design is reinforced through the specifics of blockchain technology. Pseudonymous address does not reveal the agent's identities, such that it is not possible to attach a real person to an address. For example, due to the programmability of smart contracts and interoperability of blockchain applications voting power can be "borrowed" and voted in the same transaction.

## 6      Conclusion

Decentralized finance aims to create a trust-minimized and automated version of traditional financial infrastructure. Practioneers and research alike attribute great potential to blockchain-based finance. Transparent and decentralized governance for these core protocols is of utmost importance, reinforc-ing the call for design-driven research of on-chain governance mechanisms. In this paper, we provide a measure to quantify the political governance distribution in the fast-growing ecosystem of blockchain-based finance (DeFi). Based on an empirical analysis we collect data from the Ethereum blockchain and show how the initial distribution methodologies for governance tokens may exercise significant impact on the medium-term concentration of voting power. The token distribution for all four observed projects is relatively concentrated. However, the governance mechanisms of DeFi applications poses additional high barriers for protocol changes. This results in a double-digit Nakamoto coefficient for each observed protocol.

The creation of a measurement potentially supports further IS scholars for an objective evaluation of the capabilities and limitations of on-chain token governance and distribution mechanism design. By collecting and analysing empirical data on the token economies emerging around decentralized financial applications, we aim to contribute to the growing IS discourse on the capacity for permissionless blockchain technology to enact socio-economic change through on-chain governance.

Due to the novelty of decentralized finance, the presented research is work-in-progress opening avenues for future research. First, token distributions vary over time due to economic incentives and new issuance schemes. Applying the provided framework, we want to measure a time-series of the governance power in DeFi applications. Second, our framework can be extended by not only considering objective voting power, but further analysing soft opinion building by analysing discussions and sentiment.

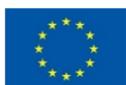 This project has received funding from the European Union's Horizon 2020 research and innovation programme under the Marie Skłodowska-Curie grant agreement No 801199

## **References**

Antonopoulos, A. and Wood, D. G. (2018) *Mastering Ethereum: Building Smart Contracts and DApps*. Sebastopol, CA: O'Reilly Media.

Azouvi, S., Maller, M. and Meiklejohn, S. (2018) 'Egalitarian Society or Benevolent Dictatorship: The State of Cryptocurrency Governance', in *Financial Cryptography Workshop*, pp. 127–143. doi: 10.1007/978-3-662-58820-8_10.

Böhme, R. *et al.* (2015) 'Bitcoin: Economics, Technology, and Governance', *Journal of Economic Perspectives*, 29(2), pp. 213–238. doi: 10.1257/jep.29.2.213.

Kolb, J. *et al.* (2020) 'Core Concepts, Challenges, and Future Directions in Blockchain', *ACM Computing Surveys*, 53(1), pp. 1–39. doi: 10.1145/3366370.

Kranz, J., Nagel, E. and Yoo, Y. (2019) 'Blockchain Token Sale: Economic and Technological Foundations', *Business and Information Systems Engineering*. Springer Gabler, 61(6), pp. 745–753. doi: 10.1007/s12599-019-00598-z.

Lindman, J., Tuunainen, V. K. and Rossi, M. (2017) 'Opportunities and Risks of Blockchain Technologies: A Research Agenda', in *Proceedings of the 50th Hawaii International Conference on System Sciences (2017)*. Hawaii International Conference on System Sciences. doi: 10.24251/hicss.2017.185.

Rossi, M. *et al.* (2019) 'Blockchain Research in Information Systems: Current Trends and an inclusive Future Research Agenda', *Journal of the Association for Information Systems*, 20(9), pp. 1388–1403. doi: 10.17705/1jais.00571.

Srinivasan, B. S. and Lee, L. (2017) *Quantifying Decentralization*. Available at: https://news.earn.com/quantifying-decentralization-e39db233c28e (Accessed: 25 October 2020).

Swan, M. (2015) *Blockchain: Blueprint for a New Economy*. 1st edn. Sebastopol, CA: O'Reilly Media, Inc.

Treiblmaier, H. (2019) 'Toward More Rigorous Blockchain Research: Recommendations for Writing Blockchain Case Studies', *Frontiers in Blockchain*. Frontiers Media SA, 2(3), p. 3. doi: 10.3389/fbloc.2019.00003.

Wright, A. and De Filippi, P. (2015) 'Decentralized Blockchain Technology and the Rise of Lex Cryptographia', *SSRN Electronic Journal*. doi: 10.2139/ssrn.2580664.

Wu, K. *et al.* (2020) 'A Coefficient of Variation Method to Measure the Extents of Decentralization for Bitcoin and Ethereum Networks', *International Journal of Network Security*, 22(2), pp. 191–200. doi: 10.6633/IJNS.202003.

Zheng, Z. *et al.* (2017) 'An Overview of Blockchain Technology: Architecture, Consensus, and Future Trends', in *Proceedings - 2017 IEEE 6th International Congress on Big Data, BigData Congress 2017*. Institute of Electrical and Electronics Engineers Inc., pp. 557–564. doi: 10.1109/BigDataCongress.2017.85.

Zoltu, M. (2019) *How to turn $20M into $340M in 15 seconds*. Available at: https://medium.com/coinmonks/how-to-turn-20m-into-340m-in-15-seconds-48d161a42311 (Accessed: 31 October 2020).